\documentclass[aps,amsmath,amssymb]{revtex4}
\usepackage{graphicx}
\usepackage{latexsym,amssymb,amsfonts,wasysym,color}
\usepackage[all]{xy}

\begin{document}
\title{On the Quantum Kinetic Approach and the
Scattering Approach to Vacuum Pair Production}
\author{Cesim K. Dumlu}
\affiliation{Department of Physics\\
University of Connecticut\\
Storrs, CT 06269-3046, USA\\
E-mail:dumlu@phys.uconn.edu
}
\begin{abstract}
For Schwinger pair production with background electric fields varying only in time, it is shown that the quantum kinetic approach, based on the quantum Vlasov equation, and the quantum mechanical scattering approach, are completely equivalent. This different perspective on the quantum kinetic approach suggests new approximation methods, which are tested in detail using a soluble pulse-like field configuration, and may give useful insight for generalization to more general electric fields, varying in more than 1 dimension.
\end{abstract}
\maketitle
\section{Introduction}
It is well known that the quantum loop corrections to the classical Maxwell Lagrangian lead to the nonlinear interactions of light due to vacuum polarization \cite{heisen,schwinger}. Some of these effects such as the Casimir effect and nonlinear Compton scattering have been experimentally observed. Others such as vacuum birefringence and electron-positron pair production from vacuum have still to be directly observed. Definite plans exist for such observations \cite{elih,elig}. At planned laser facilities, such as the Extreme Light Infrastructure (ELI), extreme field strengths might be reached and the phenomenon of pair creation in an external field could be observed \cite{eli}.\\\\
The crucial goal for theory is to find a quantitative relationship between the applied electromagnetic field and the rate at which vacuum pair creation occurs. The problem formally can be worked out using quantum field theory as the pair production rate can be deduced from the imaginary part of the effective action \cite{schwinger}. However clear explicit results exist only when the electric field is a function of a single coordinate variable. If this is the case, more straighforward approaches exist to find the pair production rate. For example by representing the laser field as a time dependent but spatially uniform electric field, one way is to reduce the time evolution to a 1 dimensional scattering problem
\cite{breyzin,popov1}, which can then be solved using the numerical or the WKB methods \cite{breyzin,popov1,kim1,kim2,kim3}. Another is the quantum kinetic approach in which the adiabatic particle number satisfies a quantum Vlasov equation \cite{kluger1,kluger2,kluger3,alkofer,blaschke}.\\\\
In this article, we show that these two rather different-looking approaches are in fact equivalent. This new perspective leads to new and improved approximation methods.
\section{Methods and Equivalence}
\subsection{Quantum Kinetic Approach}
In the quantum kinetic approach, the adiabatic particle number operator for scalar and spinor QED is defined in a time dependent basis to realize the effect of the time dependent background field \cite{kluger1,kluger2,kluger3,alkofer,thesis,blaschke}. The adiabatic particle number is then defined as the expectation number of this operator:
\begin{equation}
\mathcal{N}_{\bf{k}}(t)=\langle a^{\dagger}_{\bf{k}}(t)a_{\bf{k}}(t)\rangle=\langle b^{\dagger}_{\bf{-k}}(t)b_{\bf{-k}}(t)\rangle,
\end{equation}
where $a_{\bf{-k}}$(t)($a_{\bf{-k}}^{\dagger}(t)$) and  $b_{\bf{-k}}(t)$($b_{\bf{-k}}^{\dagger}(t)$) correspond to the annihilation (creation) operators for the particle and the antiparticle respectively with the momentum $|\bf{k}|$. The expectation value can be labeled by momentum since the background electric field is homogeneous in space. To specify the time dependent basis for the creation and annihilation operators, for scalar QED, one starts with the Klein Gordon (KG) equation. Using the Fourier decomposition of the field operator, $\Phi(\vec{x},t)=\int e^{i\vec{k}.\vec{x}}\left(u_{\bf{k}}(t)a_{\bf{k}}+u^{*}_{\bf{k}}(t)b^{\dagger}_{-\bf{k}}\right)$, and the fact that the gauge field varies only in time, the Klein Gordon equation reduces to:
\begin{equation}
\frac{d^2u_{\bf{k}}(t)}{dt^2} + \omega_{\bf{k}}^2(t)u_{\bf{k}}(t)=0.
\end{equation}
Motivated by the constant electric field case, one can make the adiabatic ansatz for $u_{\bf{k}}(t)$:
\begin{eqnarray}
&&u_{\bf{k}}(t)=\frac{\alpha_{\bf{k}}(t)}{\sqrt{2\omega_{\bf{k}}(t)}}e^{-i\phi_{\bf{k}}(t)}+\frac{\beta_{\bf{k}}(t)}{\sqrt{2\omega_{\bf{k}}(t)}}e^{i\phi_{\bf{k}}(t)}\\
&&\dot{u}_{\bf{k}}(t)=-i\omega_{\bf{k}}(t)\left(\frac{\alpha_{\bf{k}}(t)}{\sqrt{2\omega_{\bf{k}}(t)}}e^{-i\phi_{\bf{k}}(t)}-\frac{\beta_{\bf{k}}(t)}{\sqrt{2\omega_{\bf{k}}(t)}}e^{i\phi_{\bf{k}}(t)}\right).\nonumber
\end{eqnarray}
where $\alpha_{\bf{k}}(t)$ and $\beta_{\bf{k}}(t)$ are the Bogoliubov coefficients. Using the gauge $\dot{A}(t)=-E(t)$, the functions $\omega_{\bf{k}}(t)$ and $\phi_{\bf{k}}(t)$ are defined as:
\begin{eqnarray}
&&\omega_{\bf{k}}(t)=\sqrt{\epsilon_{\perp}^2+k^2_{\parallel}(t)},\hspace{5mm}\phi_{\bf{k}}(t)=\int_{-\infty}^{t}\omega_{\bf{k}}(t')dt',\\
&&k_{\parallel}(t)=k_{\parallel}-\frac{q}{c}A(t),\hspace{5mm}\epsilon_{\perp}=\sqrt{m^2+k_{\perp}^2},
\end{eqnarray}
where $k_{\parallel}$ is the momentum along the direction of the applied electric field. Consistency of the ansatz (3) leads to a coupled set of differential equations for the Bogoliubov coefficients:
\begin{eqnarray}
\dot{\alpha}_{\bf{k}}(t)=\frac{\dot{\omega}_{\bf{k}}(t)}{2\omega_{\bf{k}}(t)}\beta_{\bf{k}}(t)e^{2i\phi_{\bf{k}}(t)}, \hspace{5mm}\dot{\beta}_{\bf{k}}(t)=\frac{\dot{\omega}_{\bf{k}}(t)}{2\omega_{\bf{k}}(t)}\alpha_{\bf{k}}(t)e^{-2i\phi_{\bf{k}}(t)}.
\end{eqnarray}
A similar decomposition exists for the spinor QED, based on the Dirac equation rather than the KG equation. In a suitable spinor basis the time dependence of the field operator can be expressed again in terms of a scalar coefficient function $u_{\bf{k}}(t)$ satisfying \cite{thesis}:
\begin{equation}
\frac{d^2u_{\bf{k}}(t)}{dt^2} + (\omega^2_{\bf{k}}(t)+i\dot{k}_{\parallel}(t))u_{\bf{k}}(t)=0.
\end{equation}
Making the adiabatic ansatz:
\begin{eqnarray}
&&u_{\bf{k}}(t)=\frac{\alpha_{\bf{k}}(t)}{\sqrt{2\omega_{\bf{k}}(t)(\omega_{\bf{k}}(t)-k_{\parallel}(t))}}e^{-i\phi_{\bf{k}}(t)}+\frac{\beta_{\bf{k}}(t)}{\sqrt{2\omega_{\bf{k}}(t)(\omega_{\bf{k}}(t)-k_{\parallel}(t))}}e^{i\phi_{\bf{k}}(t)},
\\
&&\dot{u}_{\bf{k}}(t)=-i\omega_{\bf{k}}(t)\left(\frac{\alpha_{\bf{k}}(t)}{\sqrt{2\omega_{\bf{k}}(t)(\omega_{\bf{k}}(t)-k_{\parallel}(t))}}e^{-i\phi_{\bf{k}}(t)}-\frac{\beta_{\bf{k}}(t)}{\sqrt{2\omega_{\bf{k}}(t)(\omega_{\bf{k}}(t)-k_{\parallel}(t))}}e^{i\phi_{\bf{k}}(t)}\right),\nonumber
\end{eqnarray}
leads to the coupled differential equations for the Bogoliubov coefficients:
\begin{eqnarray}
\dot{\alpha}_{\bf{k}}(t)=\frac{\dot{\omega}_{\bf{k}}(t)\epsilon_{\perp}}{2\omega(t)_{\bf{k}}k_{\parallel}(t)}\beta_{\bf{k}}(t)e^{2i\phi_{\bf{k}}(t)}, \hspace{5mm}\dot{\beta}_{\bf{k}}(t)=-\frac{\dot{\omega}_{\bf{k}}(t)\epsilon_{\perp}}{2\omega_{\bf{k}}(t)k_{\parallel}(t)}\alpha_{\bf{k}}(t)e^{-2i\phi_{\bf{k}}(t)}.
\end{eqnarray}
The transformation between the time dependent operators ($a_{\bf{k}}(t)$, $b_{\bf{k}}(t)$) and the time independent operators ($\tilde{a}$$_{\bf{k}}$, $\tilde{b}$$_{\bf{k}}$) is expressed in terms of the Bogoliubov coefficients $\alpha_{\bf{k}}(t)$ and $\beta_{\bf{k}}(t)$:
\begin{equation}
\left(
                \begin{array}{cc}
                  \alpha_{\bf{k}}(t) & \pm\beta_{\bf{k}}^*(t) \\
                  \beta_{\bf{k}}(t) & \alpha^*_{\bf{k}}(t) \\
                \end{array}
              \right)\left(
                       \begin{array}{c}
                         \tilde{a}_{\bf{k}} \\
                         \tilde{b}^{\dagger}_{\bf{-k}} \\
                       \end{array}
                     \right)=\left(
                       \begin{array}{c}
                         a_{\bf{k}}(t) \\
                         b^{\dagger}_{\bf{-k}}(t)\\
                       \end{array}
                     \right).
\end{equation}
Here the upper/lower sign refers to scalar/spinor QED, respectively. Imposing the field operator commutation relations, the relationship between the Bogoliubov coefficients can be given for scalar (upper) and spinor (lower) QED respectively:
\begin{eqnarray}
|\alpha_{\bf{k}}(t)|^2 \mp |\beta_{\bf{k}}(t)|^2=1.
\end{eqnarray}
Then, for both bosons and fermions, one finds:
\begin{equation}
\mathcal{N}_{\bf{k}}(t)=|\beta_{\bf{k}}(t)|^2.
\end{equation}
The number of pairs produced with the momentum  $\bf{k}$  is given by the asymptotic value $\mathcal{N}_{\bf{k}}(\infty)$. The quantum kinetic equation for $\mathcal{N}_{\bf{k}}(t)$ follows from the time evolution of the Bogoliubov coefficients in (6) and (9).
We can express the equations (6) and (9) as  (from now on, the subscript $\bf{k}$ is suppressed for notational simplicity):
\begin{equation}
\dot{\alpha}_{\pm}(t)=\frac{W_{\pm}(t)}{2}\beta_{\pm}(t)e^{2i\phi(t)},
\hspace{5mm}\dot{\beta}_{\pm}(t)=\frac{\pm W_{\pm}(t)}{2}\alpha_{\pm}(t)e^{-2i\phi(t)},
\end{equation}
where the functions $W_{\pm}(t)$ (for scalar and spinor QED) are defined as:
\begin{equation}
W_{+}(t)=\frac{\dot{\omega}(t)}{\omega(t)}, \hspace{5mm}
W_{-}(t)=\frac{\dot{\omega}(t)\epsilon_{\perp}}{\omega(t)k_{\parallel}(t)}.
\end{equation}
Taking the time derivative of (12), and using (13), one obtains the equation for the evolution of the adiabatic particle number. Further with the aid of (11), one gets the quantum kinetic equation which is also known as the quantum Vlasov equation \cite{kluger1,kluger2,kluger3,alkofer,thesis}:
\begin{equation}
\dot{\mathcal{N}}_{\pm}(t)=\frac{W_{\pm}(t')}{2}\int_{-\infty}^{t}W_{\pm}(t)
(1 \pm 2\mathcal{N}_{\pm}(t'))\cos\left({2\int_{t'}^{t}d\tau\omega(\tau)d\tau}\right) dt'.
\end{equation}
The important physical characteristic of the quantum kinetic equations is their non-Markovian nature. The change in the particle number at a moment $t$ depends on the particle number's past configuration. In other words, the integral incorporates memory effects.
\subsection{Scattering Picture}
Another common approach to the vacuum pair production problem when the electric field just varies in time, is based on a WKB approximation to an effective one dimensional quantum mechanical scattering problem. In this approach \cite{keldysh,popov1,popov2,breyzin}, equation (2) is interpreted as a one-dimensional Schr\"{o}dinger equation with $t$ playing the role of $x$, and with ``potential'' given by $\omega^2(t)$. The physical boundary conditions for vacuum pair production correspond to the scattering boundary conditions, and the reflection coefficient can be related to the adiabatic particle number. Often this reflection coefficient is computed using the WKB approximation, but it can just as well be computed exactly (numerically). For example for scalar QED, Popov proposed the ansatz \cite{popov1}:
\begin{eqnarray}
&&u(t)=B(t)(e^{-i\phi(t)}-R(t)e^{i\phi(t)}),\\
&&\dot{u}(t)=i\omega(t)B(t)(-e^{-i\phi(t)}-R(t)e^{i\phi(t)})\nonumber.
\end{eqnarray}
The second line requires $B(t)$ and $R(t)$ to be related through:
\begin{equation}
\dot{B}(t)(e^{-i\phi(t)}+R(t)e^{i\phi(t)})+\dot{R}(t)B(t)e^{i\phi(t)}=0.
\end{equation}
This ansatz converts the ``Schr\"{o}dinger'' equation (2) into a Riccati-type equation for $R(t)$
\begin{equation}
\dot{R}(t) = \frac{\omega(t)}{2\dot{\omega}(t)}\left(e^{-2i\phi(t)} - R^2(t)e^{2i\phi(t)}\right),
\end{equation}
with initial coefficient condition $R(-\infty)=0$. Comparing (16) and (18) with (3) and (9) respectively, we identify $R(t)=\frac{\beta(t)}{\alpha(t)}$, and $|R(\infty)|^2$ corresponds to the reflection coefficient. Thus,
\begin{equation}
|R(\infty)|^2=\frac{|\beta(\infty)|^2}{|\alpha(\infty)|^2}=\frac{\mathcal{N}(\infty)}{1+\mathcal{N}(\infty)}
\end{equation}
A similar argument for spinor QED, based on (8) and (9) can be given. Together with scalar case the Riccati equations are given as:
\begin{equation}
\dot{R}_{\pm}(t) = \frac{\pm W_{\pm}(t)}{2}\left(e^{-2i\phi(t)} \mp R_{\pm}^2(t)e^{2i\phi(t)}\right), \hspace{5mm}
|R_{\pm}(\infty)|^2=\frac{\mathcal{N}_{\pm}(\infty)}{1\pm\mathcal{N}_{\pm}(\infty)}
\end{equation}
This argument shows that the quantum kinetic theory evolution equations (15) are equivalent to the 1 dimensional scattering Riccati equations (20). This is not immediately obvious looking at (15) and (20). To see this equivalance more explicitly, we note that the Riccati equation (20) involves both the amplitude and the phase of the reflection amplitude $R(t)$, but only the amplitude $|R(\infty)|^2$ is needed. We can eliminate the phase dependence to obtain second order nonlinear differential equations for $|R(t)|^2$, which are naturally expressed as:
\begin{eqnarray}
\ddot{Q}_{\pm}(t)&=&\frac{\dot{W}_{\pm}(t)}{W_{\pm}(t)}\dot{Q}_{\pm}(t) \pm W_{\pm}^{2}(t)Q_{\pm}(t)
-2\omega(t)W_{\pm}(t)\sqrt{\pm(Q_{\pm}^{2}(t)-1)-\left(W_{\pm}\dot{Q}_{\pm}(t)\right)^2},
\end{eqnarray}
where we have defined
\begin{eqnarray}
Q_{\pm}(t)&=&\frac{1\mp|R_{\pm}(t)|^2}{1\pm|R_{\pm}(t)|^2}=1\pm2\mathcal{N}_{\pm}(t).
\end{eqnarray}
The equivalence of the Riccati equation (18) to (21), for scalar QED, can be shown by defining $R_{+}(t)$ as:
\begin{equation}
R_{+}(t) = M(t)e^{2i\chi(t)},
\end{equation}
where both $M$ and $\chi$ are real. Time derivative of (23), with the aid of equation (18), yields a set of coupled differential equations:
\begin{eqnarray}
\frac{\dot{M}(t)}{1-M^2(t)}&=&\frac{W_{+}(t)}{2}\cos{\left(2\phi(t)+2\chi(t)\right)},\\
\frac{2\dot{\chi}(t)M}{1+M^2(t)}&=&-\frac{W_{+}(t)}{2}\sin{\left(2\phi(t)+2\chi(t)\right)}.
\end{eqnarray}
These coupled differential equations can be shown to equivalent to (21) using the definition (22) of $Q(t)$. To show that the quantum Vlasov equation (15) is also equivalent to (21), we differentiate (15) and use definition (22) to obtain:
\begin{eqnarray}
\ddot{Q}_{+}(t)&=&\frac{\dot{W}_{+}(t)}{W_{+}(t)}\dot{Q}_{+}(t) +W_{+}^{2}Q_{+}(t)
-2\dot{\omega}(t)\int_{-\infty}^{t} W_{+}(t')Q_{+}(t')\sin{\left(2\phi(t)-2\phi(t')\right)}dt',
\end{eqnarray}
We recognize the first two terms in (21) immediately. The final, square root, term in (21) by using (24) and (25) can be given as:
\begin{equation}
2\dot{\omega}(t)\sqrt{Q_{+}^{2}(t)-1-\left(\dot{W}_{+}(t)\dot{Q}_{+}(t)\right)^2} =\frac{4M(t)\dot{\omega}(t)}{1-M^2(t)}\sin{\left(2\phi(t)+2\chi(t)\right)}.
\end{equation}
To show the complete equivalence, setting last term of (26) equal to the righthand side of (27) requires:
\begin{eqnarray}
W_{+}(t)Q_{+}(t)\cos{(2\phi(t))}&=&\frac{d\left(\cos{2\chi(t)}\sqrt{Q_{+}^{2}(t)-1}\right)}{dt},\\
W_{+}(t)Q_{+}(t)\sin{(2\phi(t))}&=&-\frac{d\left(\sin{2\chi(t)}\sqrt{Q_{+}^{2}(t)-1}\right)}{dt}.
\end{eqnarray}
Above equations can be shown to be true after differentiating the righthand sides and utilizing (24) and (25) to eliminate $\chi(t)$. A similar argument can also  be applied for the spinor case. In these differential equations the memory effects of the Vlasov equations (15) are registered in the nonlinear square root terms in (21).\\\\
\section{Approximations}
For realistic potentials that might realize pair creation, the Schr\"{o}dinger equations (2) and (7) cannot be solved analytically. This fact makes it clear that numerical methods have to be used. Also note that the quantum kinetic approach gives $\mathcal{N}_{\bf{k}}(\infty)$, the number of particles of momentum $\bf{k}$, and the scattering approach gives $|R_{\bf{k}}(\infty)|^2$, which determines $\mathcal{N}_{\bf{k}}(\infty)$. Therefore, to obtain the total particle number one has to integrate the numerical results over the momenta, which is computationally intensive. Hence, the efficiency of these two equivalent approaches is vital.\\\\ Several approximations can be made to improve efficiency without giving away too much from accuracy.
Two basic approximations that can be done to Vlasov equations are the Low Density Approximation (LDA) and the Markovian Approximation (MA) \cite{thesis}. It is important to stress that these approximation methods are peculiar to the quantum kinetic approach. In the LDA, it is assumed that the particle density $\mathcal{N}_{\pm}(t)$ is very small. Then the  quantum kinetic equations reduce to:
\begin{equation}
\dot{\mathcal{N}}_{\pm}^{LDA}(t)=\frac{W_{\pm}(t)}{2}\int_{-\infty}^{t}W_{\pm}(t')
\cos{\left(2\int_{t'}^{t}d\tau\omega(\tau)d\tau\right)} dt'.
\end{equation}
Another approximation is the MA, in which one neglects the memory effects induced on the particle number, so that the factor $(1\pm2N_{\pm}(t))$ can be taken out of the integral, yielding:
\begin{equation}
\dot{\mathcal{N}}_{\pm}^{MA}(t)=\frac{W_{\pm}(t)}{2}(1\pm2\mathcal{N}_{\pm}^{MA}(t))\int_{-\infty}^{t}W_{\pm}(t')
\cos{\left(2\int_{t'}^{t}d\tau\omega(\tau)d\tau\right)} dt'.
\end{equation}
These two approximations are clearly related to one another by:
\begin{equation}
\dot{\mathcal{N}}_{\pm}^{MA}(t)=(1\pm2\mathcal{N}_{\pm}^{MA}(t))\dot{\mathcal{N}}_{\pm}^{LDA}(t).
\end{equation}
The LDA is physically relevant when the intensity of the applied field is very low, since the pair production rate for low intensity fields is small. This type of weak field $\left(qE(t)\ll\epsilon^2_{\perp}\right)$ also ensures the applicability of the MA. This becomes more evident, if one classifies the relevant timescales for the pair production process. The characteristic timescale for particles is expected to be order
$\tau_{particle}\sim\frac{1}{\epsilon_{\perp}}$, while the relevant time for pair production can be taken as $\tau_{pair}\sim\frac{\epsilon_{\perp}}{qE(t)}$. Then from the weak field condition above, one has:
\begin{equation}
\tau_{particle}\ll\tau_{pair}.
\end{equation}
The non-Markovian behaviour states that the number density configuration at time $t$ depends on the external field's past configuration up to time $t$.  The relevant timescale for particles is the measure of the time required to induce memory effects on the number distribution. If this time is very small compared to the relevant time scale of pair production, which depends on the energy of the field, then non-Markovian effects cannot be resolved by the system. Therefore the above inequality ensures the Markovian assumption can be applied. However, this does not mean that the system becomes completely Markovian, since there will still be  memory effects represented by the integral in (15), even if the $(1\pm2\mathcal{N}_{\pm}(t))$ term is taken out.\\\\
The equivalent quantum mechanical scattering interpretation suggests other approximations. The most obvious one is to make a WKB approximation to the scattering problem. Since the ``potential'' $-\omega^2(t)$ is negative this is an ``over-the-barrier'' scattering problem, which is conveniently described using the ``imaginary time'' method. Defining $x'=it$ we find
\begin{equation}
-\frac{d^2u(x')}{dx'^2} + \omega^2(x')u(x')=0 ,
\end{equation}
so we have a one dimensional quantum mechanical tunneling problem. In this case, the asymptotic value of the particle number corresponds to the transmission or tunneling probability. Identifying the problem as a tunneling process enables us to make the WKB approximation. In WKB approximation, the tunneling amplitude is given by \cite{kim3}:
\begin{equation}
|R(\infty)|^2\approx e^{-2S_{\mathrm{classical}}},\hspace{5mm}S_{\mathrm{classical}}=\oint_{C}\mathcal{L}(t)dt,
\end{equation}
where $S_{\mathrm{classical}}$ is the classical action. The WKB approximation is valid in the quasiclassical regime in which the following inequalities hold:
\begin{equation}
E\ll m^2,\hspace{5mm}\frac{1}{\tau}\ll m,\hspace{5mm}\frac{1}{\tau^2}\ll E.
\end{equation}
Here $\tau$ stands for a characteristic width of the time dependence of the  external field, and $E$ is the peak magnitude. The  WKB approximation requires the computation of the contour integral expression for the classical action.\\\\
Returning to the Riccati equation (20), a simpler approximation method can be introduced. From (19) it can inferred that in the LDA the limit $\mathcal{N}_{\pm}(t)\rightarrow0$ implies taking the limit $R_{\pm}(t)\rightarrow0$. Then, the Riccati equation for scalar and spinor cases can be reduced to:
\begin{equation}
\dot{R}_{\pm}(t) =\frac{\pm W_{\pm}(t)}{2}e^{-2i\phi(t)},
\end{equation}
neglecting the nonlinear term. We call this the Truncated Riccati Approximation (TRA).
Although this approach seems to be  mathematically  equivalent to the LDA, its numerical evaluation yields almost the same results with that of the MA.\\\\
A comparison between these various approximations provides valuable insight to the pair production problem. This can be done for a pulse shaped time dependent potential for which the exact solution is available. Therefore, a comprehensive assessment can be done on the basis of accuracy and efficiency.
\section{Comparison}
In this section we compare the mentioned approximation methods using an exactly soluble pulse-shaped field:
\begin{equation}
A(t)=-E\tau\tanh{\frac{t}{\tau}},\hspace{5mm}E(t)=\frac{E}{(\cosh{\frac{t}{\tau}})^2},
\end{equation}
for which the KG and Dirac equations are exactly soluble \cite{niki}. Therefore, a closed analytic form of the particle number, for both scalar and spinor QED; can be given respectively as \cite{popov1,niki}:
\begin{eqnarray}
|R_{+}(\infty)|^2&=&\frac{\cosh{(\alpha-\beta)}+\cos{\lambda}}{\cosh{(\alpha+\beta)}+\cos{\lambda}},\\
|R_{-}(\infty)|^2&=&\frac{\sinh{(\frac{\alpha-\beta+\lambda'}{2})}+\sinh{(\frac{\alpha-\beta-\lambda'}{2}})}{\sinh{(\frac{\alpha+\beta+\lambda'}{2})}+\sinh{(\frac{\alpha+\beta-\lambda'}{2})}}.
\end{eqnarray}
The parameters are defined as:
\begin{eqnarray}
\alpha=\pi\tau\sqrt{m^2+k^2_{\perp}+(k_{\parallel}+eE)^2},& \beta=\pi\tau\sqrt{m^2+k^2_{\perp}+(k_{\parallel}-eE)^2},\\
\noindent\lambda=\pi\tau\sqrt{\frac{1}{\tau^2}-(2eE\tau)^2},&  \lambda=i\lambda'.&
\end{eqnarray}
The exact closed form of the particle number becomes useful to check for computational errors and the validity of the numerical methods. For both scalar  and spinor cases the agreement between the Riccati, quantum kinetic equations and the closed form is well established up to 7-8 significant figures in Mathematica for general values of the parameters. Here we present the results of the comparison between the approximation schemes in the relevant domains of momentum and electric field intensity.
\begin{figure}
\begin{picture}(1,1)
\put(189,140){\textcolor[rgb]{0.00,0.00,1.00}{.....}}
\put(204,139){\tiny{WKB}}
\put(190,130){\textcolor[rgb]{0.65,0.33,0.33}{-$\cdot$-$\cdot$}}
\put(204,129){\tiny{LDA, MA ,TRA}}
\put(190,120){\textcolor[rgb]{0.52,0.52,0.26}{\line(35,0){11}}}
\put(204,119){\tiny{Exact}}
\end{picture}
\includegraphics[width=4in]{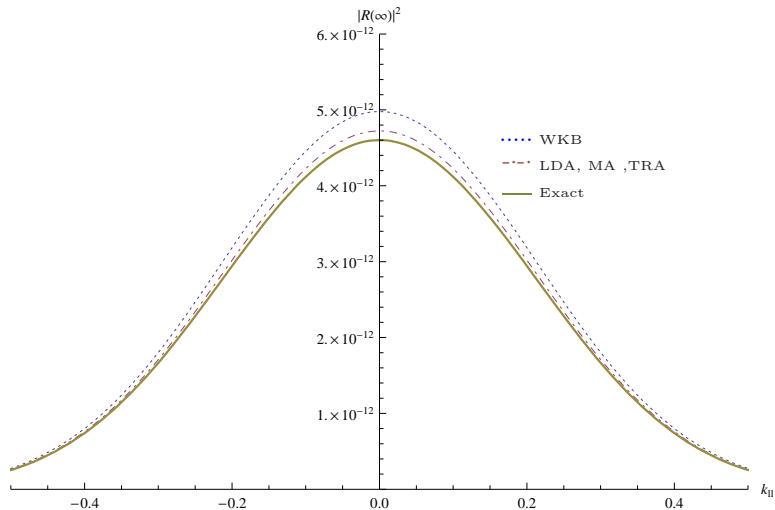}
\caption{Quasiclassical regime, scalar case. Four approximation schemes are indicated. The plot shows the bell-shaped dependence of pair production rate on $k_{\parallel}$, in units m. The field intesity is taken as $E=0.1E_{cr}=0.1m^2$, and the field width is set to $\tau=10m$. Mass and charge are taken as 1. Normalized units ($\hbar=1, c=1$) were used for all numerical calculations.}
\end{figure}
\begin{figure}
\begin{center}
\begin{picture}(1,1)
\put(216,150){\textcolor[rgb]{0.00,0.00,1.00}{....}}
\put(233,149){\tiny{WKB}}
\put(217,140){\textcolor[rgb]{0.65,0.33,0.33}{\line(35,0){11}}}
\put(233,139){\tiny{Exact}}
\put(217,130){\textcolor[rgb]{0.52,0.52,0.26}{-$\cdot$}\textcolor[rgb]{0.00,0.65,0.33}{-$\cdot$}}
\put(233,129){\tiny{\textcolor[rgb]{0.52,0.52,0.26}{TRA}, \textcolor[rgb]{0.00,0.65,0.33}{MA}}}
\put(217,120){\textcolor[rgb]{0.00,0.38,0.57}{- -}}
\put(233,119){\tiny{LDA}}
\end{picture}
\includegraphics[width=4in]{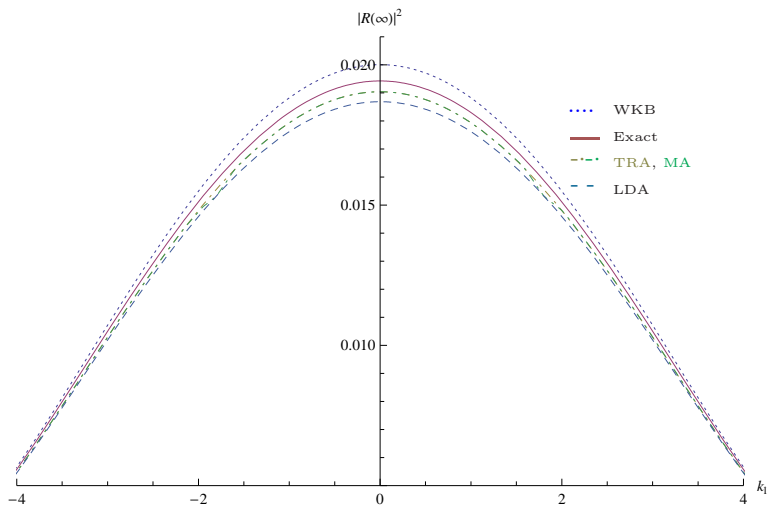}
\caption{The high intensity field regime, scalar case. Again the bell-shaped dependence of pair production rate on $k_{\parallel}$ is observed. Electric field intensity is taken as  $E = 0.8E_{cr} = 0.8m^2$. The field width is taken as $\tau=10m$. Mass and charge are taken as 1. It is observed that the rate changes drastically as the field intensity increases.}
\end{center}
\end{figure}
\subsection{Scalar QED}
The first relevant domain is the quasiclassical regime where the inequalities in (36) hold. In this regime, the  WKB approximation becomes valid and all four schemes can be compared. For the chosen gauge field, the WKB approximation yields the pair production rate as \cite{kim3}:
\begin{equation}
e^{-2S_{\mathrm{classical}}}=e^{-(\alpha+\beta-\lambda')}.
\end{equation}
In Figure 1, the approximations are compared with the exact closed form (39). The numerical results show that in the quasiclassical regime the WKB approximation overshoots the exact result when the momentum gets close to zero. The same pattern is observed for all remaining approximations. However they differ from the exact result up to $\sim\%5$. In this regime, the TRA, MA and the LDA results are indistinguishable.\\

The other relevant regime is the domain where the electric field intensity gets close to the critical field ($E=m^2$, see Figure 2.). In this domain, the TRA and the MA give closer results to the exact answer than the LDA. However results of these 3 approximations lie below the analytic result unlike, in the quasiclassical regime.\\

\subsection{Spinor QED}
In the quasiclassical regime of spinor QED, the WKB approximation gives the closest result to the analytical expression. The closed form of the WKB method is given as \cite{kim3}:
\begin{eqnarray}
e^{-2S_{\mathrm{classical}}}&=&e^{-(\alpha+\beta-\lambda')}e^{\frac{\sigma}{2}\left(\sqrt{1+i\pi\frac{2}{\sigma}}
+\sqrt{1-i\pi\frac{2}{\sigma}}-2\right)},\\
\sigma&=&2\pi eE\tau.
\end{eqnarray}
For spinor QED, the momentum distribution falls off faster than in the scalar case. However, similar to the scalar case, the TRA, MA and the LDA give results on top of each other. When the electric field intensity gets higher, the WKB approximation still gives the best result.
\begin{figure}
\begin{center}
\begin{picture}(1,1)
\put(200,140){\textcolor[rgb]{0.00,0.00,1.00}{-$\cdot$-$\cdot$}}
\put(214,139){\tiny{LDA, MA ,TRA}}
\put(200,130){\textcolor[rgb]{0.65,0.33,0.33}{\line(35,0){11}}}
\put(214,129){\tiny{Exact}}
\put(199,120){\textcolor[rgb]{0.52,0.52,0.26}{.....}}
\put(214,119){\tiny{WKB}}
\end{picture}
\includegraphics[width=4in]{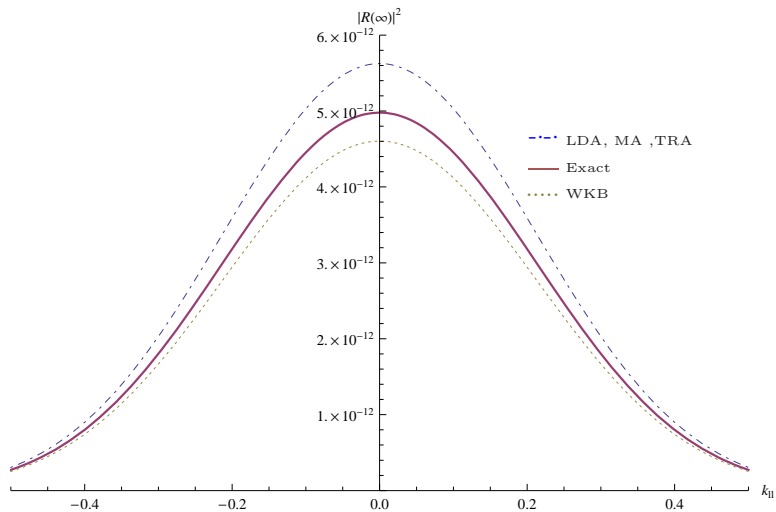}
\caption{Quasiclassical regime for the spinor case. As in the scalar QED example in Figure 1, we take  $E = 0.1E_{cr} = 0.1m^2$, and $\tau=10m$. Plot shows the values for pair production rate differs slightly than the scalar case.}
\end{center}
\end{figure}
\begin{figure}
\begin{center}
\begin{picture}(1,1)
\put(217,150){\textcolor[rgb]{0.00,0.00,1.00}{- -}}
\put(233,149){\tiny{LDA}}
\put(216,140){\textcolor[rgb]{0.52,0.52,0.26}{-$\cdot$-$\cdot$}}
\put(233,139){\tiny{TRA, MA}}
\put(216,130){\textcolor[rgb]{0.00,0.76,0.76}{\line(35,0){11}}}
\put(233,129){\tiny{Exact}}
\put(216,120){\textcolor[rgb]{0.00,0.44,0.87}{....}}
\put(233,119){\tiny{WKB}}
\end{picture}
\includegraphics[width=4in]{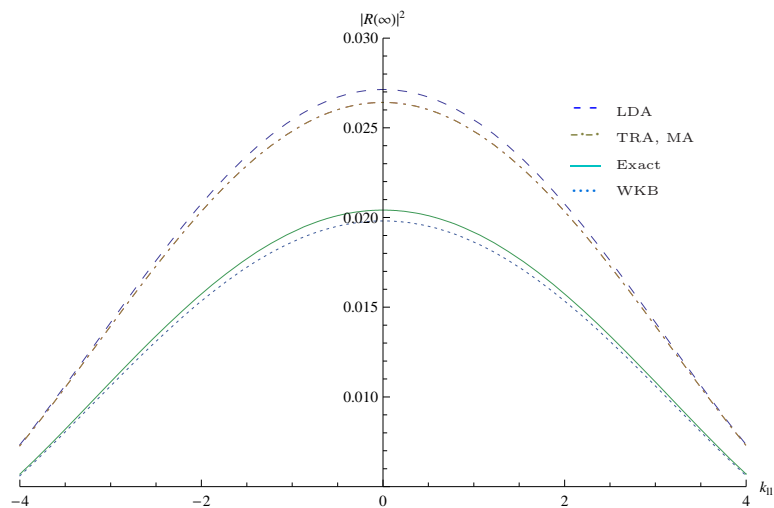}
\caption{High intensity field regime, spinor case. As in the scalar QED example in Figure 2, we take $E=0.8E_{cr}$, and $\tau=10m$.}
\end{center}
\end{figure}
Like in the scalar case, in this regime the distinction between the TRA, MA and the LDA gets clearer. The LDA differs from the other approximations by up to $\sim\%5$. The difference between the analytic results goes up to $\sim\%25$.\\
\section{Conclusion}
For intensities lower than the critical field the number of produced pairs
can be calculated within the QED effective action formalism. For external electric fields varying
only in time, the Riccati and quantum kinetic equations can be used for numerical evaluation and they have been shown to be equivalent. The scattering approach to the pair production problem gives a new perspective and suggests other approximation schemes which have been compared with the Low Density Approximation (LDA) and the Markovian Approximation (MA) of the quantum kinetic approach. The numerical data suggests that the Truncated Riccati Approximation (TRA) gives the same results as the (MA) of quantum kinetic approach. Among the discussed approximation schemes, the TRA appears to be the simplest one for numerical calculation. The particle number can also be calculated by using the second order nonlinear differential equations featuring the non-Markovian behaviour. However, the efficiency of these methods requires further investigation.\\\\ Although these two methods are equivalent, they can have different physical interpretations. In the kinetic approach,  the pair production rate is physically related to the particle number. On the other hand, using the Riccati equations, the pair production rate can be associated with the reflection coefficient. Another physical interpretation can be made on the basis that (2), in fact, represents a harmonic oscillator equation with a variable frequency. Then the problem of pair creation for the scalar particles becomes equivalent to the parametric excitation of an oscillator \cite{popov2}. In this respect, the excitation parameter becomes the pair production rate. As for the spinor case, the same procedure starts with the Dirac equation and in one dimension as suggested in \cite{popov1}, the problem can be viewed as a problem of spin precession. These arguments suggest that looking at the problem from different standpoints, one may calculate the rate and make approximations relevant to the particular perspective taken. However, the  applicability of the quantum kinetic approach is restricted to the fields varying in one dimension. If the scattering approach can be generalized for the fields varying in more than one dimension and for one dimensional fields with more complicated structure \cite{dunne}, it might give useful insight to the vacuum pair production problem.
\begin{acknowledgments}
CKD gratefully thanks to G. Dunne for fruitful discussions and valuable suggestions.
\end{acknowledgments}

\end{document}